\documentclass[a4paper,11pt]{article}
\usepackage{aaskaiid}
\usepackage{xcolor}
\usepackage{siunitx}
\usepackage{physics}
\usepackage{orcidlink}
\title{Using SKA-Low to Detect PeV Gamma-rays from Galactic Sources}
\ShortTitle{Using SKA-Low to detect PeV $\gamma$-rays}

\abstract{Detecting so called PeVatrons is considered one of the prime goals of $\gamma$-ray astronomy. PeVatrons are astrophysical objects in the Galaxy that are sources of cosmic rays exceeding PeV ($10^{15}$ eV) energies, the highest in our Galaxy. Their nature is unknown as of now, with some candidates reaching barely above PeV energies just having been identified. Serendipitously, the energy threshold of air shower detection using radio emission, has been proven at 50 PeV. There is a case to be made that SKA-Low with its unprecedented number of antennas, can reach lower in energy, while the size of the core is sufficiently large provide a significant effective area to measure PeV fluxes. While this promises a novel angle towards understanding the cosmic ray accelerators in our Galaxy, it also would be the first detection of $\gamma$-ray air showers using radio emission.}

\begin{document}

\author*[1,2]{Anna~Nelles\orcidlink{0000-0002-1720-6350}}
\emailAdd{anna.nelles@fau.de}
\author[1]{Philipp~Laub\orcidlink{0009-0003-2617-9109}}
\author[3,4]{Haoning~He\orcidlink{0000-0002-8941-9603}}
\author[14]{Felix~Schl\"uter\orcidlink{0000-0002-5545-4363}}

\author[1]{Sjoerd~Bouma\orcidlink{0000-0002-6959-2302}}
\author[5]{Justin~Bray\orcidlink{0000-0002-0963-0223}}
\author[6,7]{Stijn~Buitink\orcidlink{0000-0002-6177-497X}}
\author*[6,7]{Arthur~Corstanje\orcidlink{0000-0001-5992-6228}} 
\author[6]{Vital~De Henau\orcidlink{0009-0003-0337-3558}}
\author[8]{Edwin~Dickinson\orcidlink{0000-0003-0834-4708}}
\author[9,10]{Brian~Hare\orcidlink{0000-0001-5138-1235}}
\author[7,6,11]{J\"org~H\"orandel\orcidlink{0000-0001-6604-547X}}
\author[12,6]{Tim~Huege}
\author[8]{Clancy~James\orcidlink{0000-0002-6437-6176}}
\author[3]{Xingyu Li\orcidlink{0009-0003-1209-2643}}
\author[12]{Hermann-Josef~Mathes}
\author[7,11]{Katharine~Mulrey\orcidlink{0000-0001-8026-8020}}
\author[12,13]{Subhadip~Saha\orcidlink{0000-0003-2435-8317}}
\author[9]{Olaf~Scholten\orcidlink{0000-0003-3649-1254}}
\author[5]{Ralph~Spencer\orcidlink{0009-0009-6015-1787}}
\author[10]{Christopher~Sterpka\orcidlink{0000-0001-8217-0836}}
\author[1]{Karen~Terveer\orcidlink{0009-0002-9594-0419}}
\author[15]{Satyendra~Thoudam\orcidlink{0000-0002-7066-3614}}
\author[16]{Gia~Trinh\orcidlink{0000-0002-5352-5092}}
\author[10]{Paulina~Turekova\orcidlink{0009-0006-1262-7507}}
\author[12]{Darko~Veberic\orcidlink{0000-0003-2683-1526}}
\author[12]{Keito~Watanabe\orcidlink{0000-0003-0599-4035}}
\author[20]{Hiroaki~Yamamoto}
\author[17,18]{Chao~Zhang\orcidlink{0000-0001-9366-0056}}
\author[19]{Pengfei~Zhang\orcidlink{0000-0002-6855-5315}}
\author[3]{Yi~Zhang\orcidlink{0000-0001-6223-4724}}

\ShortName{A.~Nelles et al.}
\newcommand{\affilASTRON}{Netherlands Institute for Radio Astronomy (ASTRON), Dwingeloo, The Netherlands}
\newcommand{\affilCanTho}{Physics Education Department, School of Education, Can Tho University, Campus~II, 3/2 Street, Ninh Kieu District, Can Tho City, Viet Nam}
\newcommand{\affilCurtin}{International Centre for Radio Astronomy Research, Curtin University, Bentley, 6102, WA, Australia}
\newcommand{\affilDESY}{Deutsches Elektronen-Synchrotron DESY, Platanenallee~6, 15738 Zeuthen, Germany}
\newcommand{\affilErlangen}{Erlangen Centre for Astroparticle Physics, Friedrich-Alexander-Universit\"at Erlangen-N\"urnberg, 91058 Erlangen, Germany}
\newcommand{\affilGorlitz}{Deutsches Zentrum f\"ur Astrophysik, Postplatz~1, 02826 Görlitz, Germany}
\newcommand{\affilGroningen}{Kapteyn Astronomical Institute, University of Groningen, P.O.~Box 72, 9700 AB Groningen, Netherlands}
\newcommand{\affilHefei}{School of Astronomy and Space Science, University of Science and Technology of China, Hefei 230026, China}
\newcommand{\affilKanpur}{Department of Physics, Indian Institute of Technology Kanpur, Kanpur, UP-208016, India}
\newcommand{\affilKeyNanjing}{Key Laboratory of Modern Astronomy and Astrophysics, Nanjing University, Ministry of Education, Nanjing 210023, China}
\newcommand{\affilKIT}{Institut f\"ur Astroteilchenphysik, Karlsruhe Institute of Technology (KIT), P.O.~Box 3640, 76021 Karlsruhe, Germany}
\newcommand{\affilKhalifa}{Department of Physics, Khalifa University, P.O.~Box 127788, Abu Dhabi, United Arab Emirates}
\newcommand{\affilManchester}{Jodrell Bank Centre for Astrophysics, Department of Physics and Astronomy, University of Manchester, Manchester M13 9PL, UK}
\newcommand{\affilMaxPlanck}{Max-Planck Institut f\"ur Astrophysik, Karl-Schwarzschild-Str.~1, 85748 Garching, Germany}
\newcommand{\affilMunich}{Ludwig-Maximilians-Universit\"at M\"unchen (LMU), Geschwister-Scholl-Platz~1, 80539 M\"unchen, Germany}
\newcommand{\affilNanjing}{School of Astronomy and Space Science, Nanjing University, Nanjing 210023, China}
\newcommand{\affilNijmegen}{Department of Astrophysics/IMAPP, Radboud University Nijmegen, P.O.~Box 9010, 6500 GL Nijmegen, The Netherlands}
\newcommand{\affilNikhef}{Nikhef, Science Park Amsterdam, 1098 XG Amsterdam, The Netherlands}
\newcommand{\affilPurpleMt}{Key Laboratory of Dark Matter and Space Astronomy, Purple Mountain Observatory, Chinese Academy of Sciences, No.~10 Yuanhua Road, Nanjing, China}
\newcommand{\affilULB}{Universit\'e Libre de Bruxelles, Science Faculty CP230, B-1050 Brussels, Belgium}
\newcommand{\affilVUB}{Inter-University Institute For High Energies (IIHE), Vrije Universiteit Brussel (VUB), Pleinlaan 2, 1050 Brussels, Belgium}
\newcommand{\affilXidian}{School of Electronic Engineering, Xidian University, No.2 South Taibai Road, Xi'an, China}
\newcommand{\affilSKA}{SKA Observatory, Jodrell Bank, Lower Withington, Macclesfield, SK11 9FT, UK}
\newcommand{\affilIITK}{Department of Physics, Indian Institute of Technology Kanpur, Kanpur, UP-208016, India}
\newcommand{\affilJap}{Department of Physics, School of Science, Nagoya University
Furo-cho, Chikusa-ku, Nagoya, Aichi, Japan, 464-8602}

\affiliation[1]{\affilErlangen}
\affiliation[2]{\affilDESY}
\affiliation[3]{\affilPurpleMt}
\affiliation[4]{\affilHefei}
\affiliation[5]{\affilULB}
\affiliation[6]{\affilManchester}
\affiliation[7]{\affilVUB}
\affiliation[8]{\affilNijmegen}
\affiliation[9]{\affilCurtin}
\affiliation[10]{\affilGroningen}
\affiliation[11]{\affilASTRON}
\affiliation[12]{\affilNikhef}
\affiliation[13]{\affilKIT}
\affiliation[14]{\affilIITK}
\affiliation[15]{\affilKhalifa}
\affiliation[16]{\affilCanTho}
\affiliation[17]{\affilNanjing}
\affiliation[18]{\affilKeyNanjing}
\affiliation[19]{\affilXidian}
\affiliation[20]{\affilJap}




\maketitle

\section{Introduction}

Radio detection of particle induced air showers has been proven feasible in the early 2010s with the advent of fast digital electronics and radio telescopes like LOFAR \citep{LOFAR:2013jil}. While the dedicated cosmic ray experiments with comparatively few antennas (\cite{PierreAuger:2015hbf,Tunka-Rex:2016nto}, e.g.) have shown that the measurement results are in agreement with particle based or optical methods, LOFAR with its comparatively high antenna density of about 600 low-band antennas in 0.1 km$^2$ at its central core, has enabled the study of the detailed features of the emission. Hereby, from every low-band antenna the raw voltages are buffered and read out following a trigger from a particle array (cf.~\cite{SKA_Book_CR_Overview}). In offline data analysis, nanosecond-scale pulses are selected, whose amplitudes map out the so called \emph{footprint} of the air shower on the ground. A footprint is a roughly ellipsoid shaped area of a few hundred m$^2$ which is illuminated with particles and radio emission by an impinging air shower from directions well above the horizon. In comparison, the central core of SKA-Low \citep{braun2019anticipatedperformancesquarekilometre} makes LOFAR look like a very sparsely instrumented telescope, with about a factor of 100 increase in antenna density. While this incredible wealth of data will enable very detailed studies of hadronic air showers (cf.~\cite{SKA_Book_Cosmic_Rays}) or even particle physics in air showers (cf.~\cite{SKA_Book_Hadronic}), we have good evidence that sheer number of antennas will allow us to lower the energy threshold of the observations. Across all energies, and in particular in the energy range of other radio air shower experiments, the cosmic ray sky is dominated by hadronic cosmic rays. At single digit PeV energies, a sub-dominant but measurable flux of $\gamma$-ray induced showers may be expected, which would open up complementary observations of the most powerful particle accelerators in our Galaxy.

Lowering the energy threshold of the observations is envisioned to be enabled through beamforming techniques that improve the signal-to-noise ratio as compared to single antennas detections. Due to the restricted size of the air shower footprint of only a few 100s of meters, only antennas very close to each other can be beamformed without loss of spatial information. Of all radio detectors currently in existence, under construction or even planned, only SKA-Low provides a sufficient density of high-quality antennas to make such an approach worth entertaining.

This article will motivate the scientific interest in the PeV range, show simulations of the anticipated detections and summarize open questions that will have to be solved before $\gamma$-ray observations with SKA-Low can become a reality.

\section{Sources of Cosmic Rays and PeV Gamma Rays}

The quest to identify and study \emph{PeVatrons}, astrophysical sources capable of accelerating cosmic rays to PeV energies ($10^{15}$ eV), represents a central challenge in modern high-energy astrophysics \citep{deOnaWilhelmi:2024mul,Cardillo:2023hbb}. These extreme accelerators are believed to be the origin of the highest-energy cosmic rays within our Galaxy. Their detection and characterization through the corresponding $\gamma$-ray emission provide a unique window into particle acceleration processes under the most extreme conditions.

\subsection{Theoretical motivation considerations for PeVatrons}

In astrophysical sources and their magnetic fields, charged particles are accelerated that reach Earth in the form of cosmic rays. They were discovered by Victor Hess more than a century ago, and their nature and influence on star formation, structure formation in the Universe, and even the formation of life have been studied since. The highest energies cosmic rays arriving at Earth still pose a number of questions. In particular, the deflection in magnetic fields has precluded the identification of their sources in our Galaxy. Based on very fundamental particle physics, it is expected that at sites of particles acceleration $\gamma$-rays are also created, through $\pi$-production (\emph{pion-production}) or $p\gamma$-interaction, which transfers energy from the charged cosmic rays to neutral $\gamma$-rays that then retain direction information. In addition, interactions of cosmic rays with interstellar matter produce high-energy diffuse gamma rays from the Galactic disk. Cosmic rays with energies up to a few 100 PeV have gyroradii of only a few tens of parsec, which is much smaller than the kiloparsec-scale size of the diffusion halo boundary of the Galaxy. As a result, they can be effectively confined by the Galactic magnetic field. The observation of PeV $\gamma$-rays will provide valuable information about the distribution of the highest energy cosmic rays of Galactic origin.

Extragalactic sources are generally not expected to contribute significantly to the observed PeV $\gamma$-ray flux \citep{HESS:2025gpm}. This is due to the severe attenuation of PeV photons by cosmic microwave background (CMB) photons and extragalactic background light (EBL) through pair production ($\gamma+\gamma\rightarrow{e^++e^-}$). This effectively imposes an opacity horizon for PeV gamma rays, rendering most extragalactic sources undetectable at these energies and emphasizing the Galactic origin of any detected PeV emission \citep{Dominguez:2013lfa}.

\subsection{Observed source classes at the highest energies}

Recent observations, particularly from instruments such as LHAASO \citep{LHAASO:2023rpg}, HAWC \citep{Abeysekara:2017hyn}, and H.E.S.S. \citep{HESS:2018pbp}, have begun to reveal a population of Galactic sources emitting $\gamma$-rays beyond 100 TeV, with a select few even exhibiting photons above 1 PeV. 
The hunt for PeVatrons has expanded beyond traditional candidates such as supernova remnants (SNRs) to include a wider array of objects. 
Evidence was presented across microquasars and $\gamma$-ray binaries, which provide compact, powerful engines; young massive star clusters (YMCs), where collective winds and shocks may act as accelerators, and pulsars, pulsar wind nebulae (PWNe), and pulsar halos, which are increasingly recognized as significant contributors to the local cosmic-ray electrons and positrons. In the following subsections, we provide a brief description of the most plausible source classes and list potential candidates for SKA-Low. 

\subsubsection{Microquasars and Gamma-Ray Binaries}
Microquasars are Galactic binary systems consisting of a compact object, a stellar-mass black hole, or a neutron star, accreting matter from a companion star and producing relativistic jets. 
Their relativistic jets and intense environments provide conditions suitable for extreme particle acceleration. 
Recent observations from H.E.S.S., HAWC, and LHAASO \citep{HESS:2024rlh,Alfaro:2024cjd,LHAASO:2024psv}, have significantly advanced the study of these systems at the highest $\gamma$-ray energies in our Galaxy. 
Although not all microquasars visible to existing $\gamma$-ray experiments have been detected at the highest energies, they remain exciting sources, where jets and the related processes of particle acceleration can be observed in Galactic sources.
Currently, interesting candidates are SS 433, V4641 Sgr, GRS 1915 + 105, MAXI J1820 + 060, Cygnus X-1, and Cygnus X-3. 

\subsubsection{Young Massive Stellar Clusters}
Young Massive stellar Clusters (YMCs) are prime targets for PeV $\gamma$-ray observations, as they represent natural candidates for being Galactic PeVatrons. The detection of photons of up to 2.5 PeV from the Cygnus region by LHAASO identifies its YMC as a potential super-PeVatron \citep{LHAASO:2023uhj}. Morphological and spectral studies further reveal spatial correlations between very-high-energy $\gamma$-ray emission and structures traced by HI gas and molecular clouds, indicating the presence of high-energy cosmic rays. Additional evidence comes from starforming regions like W43, where emission exceeding 100 TeV has been detected \citep{LHAASO:2024lls}. These findings collectively underscore YMCs as efficient particle accelerators, making them essential objects for future PeV $\gamma$-ray detections with SKA-Low.

\subsubsection{Pulsar Wind Nebulae and TeV halos}
Pulsars, pulsar wind nebulae (PWNe), and TeV halos constitute a key target class for PeV $\gamma$-ray astronomy, serving as unique laboratories for studying cosmic-ray electron acceleration and transport. These systems exhibit evidence of particle acceleration up to PeV energies, as demonstrated by LHAASO's detection of PeV emission from the Crab Nebula \citep{LHAASO:2021cbz} and other PWNe. The growing catalog of TeV halos, such as LHAASO J1847+0925 \citep{Hu:2025SD} and HESS J1831-098 \citep{Sabri:2025qxd}, reveal suppressed and anisotropic diffusion in their vicinity, which is critical for understanding cosmic ray propagation. The diversity of PeVatron candidates, from young PWNe like the Crab Nebula to middle-aged systems like HESS J1849-000~\citep{Amenomori2023}, underscores the importance of pulsar-related systems in the quest to identify Galactic PeVatrons and understand cosmic-ray origin.

\subsubsection{Supernova Remnants}
Supernova remnants (SNRs) represent a foundational target class for PeV $\gamma$-ray astronomy, serving as primary candidates for Galactic PeVatrons, sources capable of accelerating cosmic rays to PeV energies. 
Mounting observational evidence confirms their role as efficient particle accelerators. Several SNRs, such as the W51 region (hosting both the cloud-interacting W51C and the cluster-rich W51B, \cite{LHAASO:2024fym}) and the Boomerang Nebula (G106.3 + 2.7, i.e., LHAASO J2226 + 6057, \cite{LHAASO:2021Nat,TibetASG:2021NatA,Chang:2022fvj}), exhibit $\gamma$-ray emission extending beyond 100 TeV. The detection of such high-energy photons provides direct evidence of particle acceleration to PeV energies within these remnants. Through detailed spectral and morphological studies, SNRs offer a critical window into the origin and acceleration of Galactic cosmic rays, making them essential targets for next-generation PeV $\gamma$-ray observatories.

\subsubsection{Unidentified Sources}
We highlighted a number of sources that are primary objects of interest, but at the same time not always from fully disjunct source classes. The transition between a SNR and a PWN is gradual, and it is not yet clear from where the highest $\gamma$-emission is detected. Does it stem from the acceleration source or where the accelerated cosmic rays interact with a target? This makes the association of sources often not unique (e.g.\,\cite{Mitchell:2024oue,Peron:2024lfo}). 

Furthermore, several key observations are still listed as unidentified sources, since their observed location does not coincide with an obvious Galactic object or is coincident with too many. 
Key unidentified LHAASO sources, including the promising PeVatron candidate LHAASO J2018+3651 \citep{Li:2025NF}, LHAASO J1908+06 \citep{DeSarkar:2022bnl}, and LHAASO J0341+5258, \citep{LHAASO:2021plb}, represent high-priority targets for probing the nature of Galactic PeVatrons through very-high-energy $\gamma$-ray observations.

\subsection{Potential Sources for SKA-Low}

Despite these exciting discoveries, the current number of firmly identified PeVatrons remains small. This scarcity is primarily due to the relatively small effective area and poor sensitivity of  existing $\gamma$-ray observatories at very high energies. New-generation hybrid detector arrays such as LHAASO, which combine Water Cherenkov detectors with scintillation detectors, have made groundbreaking discoveries, however their sensitivity becomes quite poor above $\sim1$~PeV. Imaging Atmospheric Cherenkov Telescopes provide superior angular resolution and excellent background rejection, but they also suffer from poor sensitivity above $\sim\,100$~TeV, and their limited field-of-view, low duty cycle and relatively small effective areas make detections of gamma rays at the very highest energies ($\gg100$ TeV) statistically challenging.

\begin{figure}
    \centering
    \includegraphics[width=0.8\linewidth]{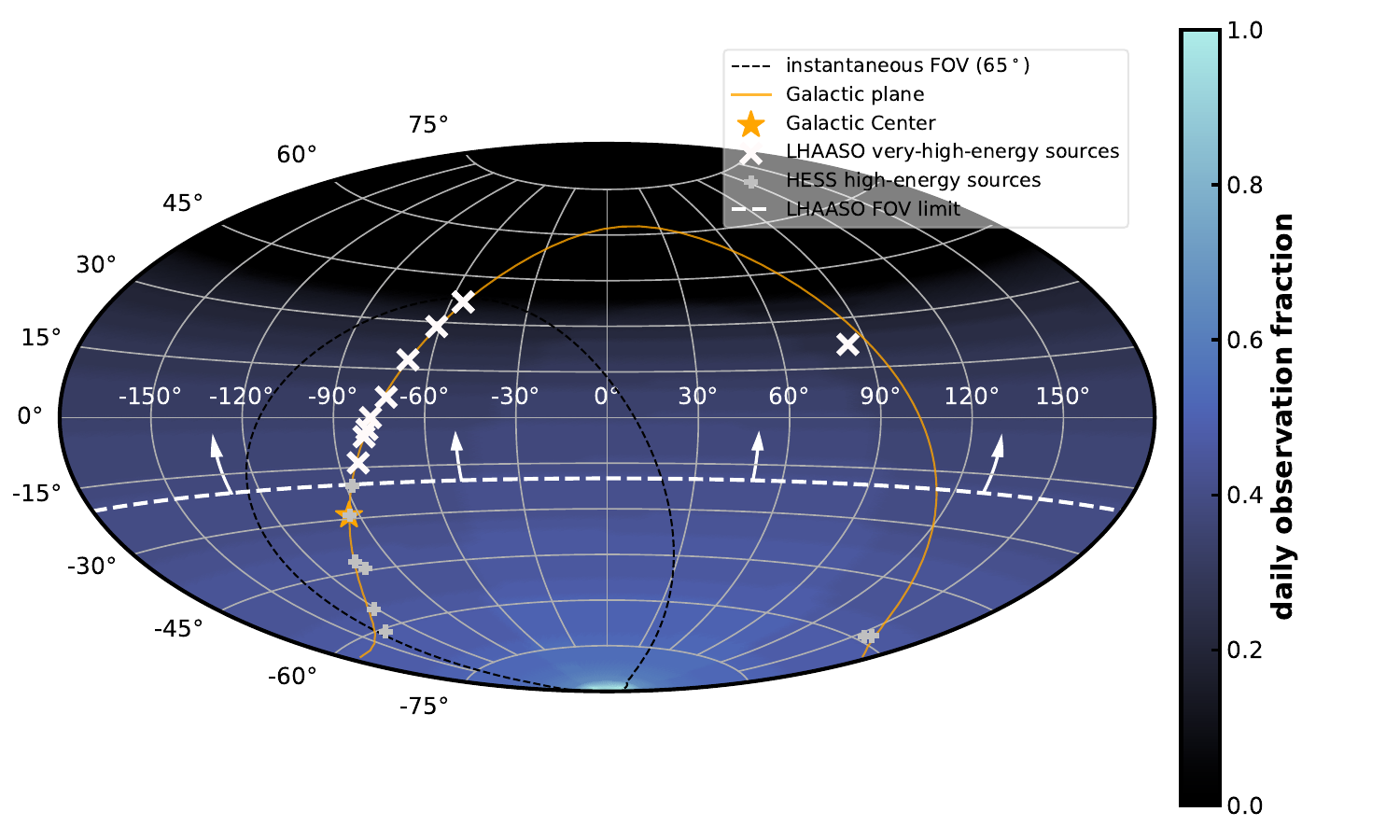}
    \caption{SKA-Low field of view to air showers in equatorial coordinates. The colorbar shows the daily observable fraction (year-averaged). Overlaid are sources of interest as detected with LHAASO \citep{LHAASO:2023rpg} and H.E.S.S \citep{HESS:2018pbp}. Figure from \cite{ICRC_2025_Felix}}
    \label{fig:skymap}
\end{figure}

This is where the potential of the SKA-Low lies. Its immense antenna array could lower the detectable energy threshold for radio-detected air showers into the crucial sub-10 PeV range and offer a large effective area precisely where traditional $\gamma$-ray instruments begin to lose sensitivity. Furthermore, its wide-field capability enables efficient detection of very-large-zenith-angle events, greatly expanding the accessible sky and observation time. Figure \ref{fig:skymap} shows an illustration of the field of view and a number of sources already discovered, as discussed above. In particular, in the field-of-view only covered by H.E.S.S. many additional sources are expected to be discovered, given the appropriate amount of observation time. LHAASO is an experiment on the Northern hemisphere, therefore essentially blind to Southern sources visible to SKA-Low. 

\begin{figure}
    \centering
    \includegraphics[width=0.6\linewidth]{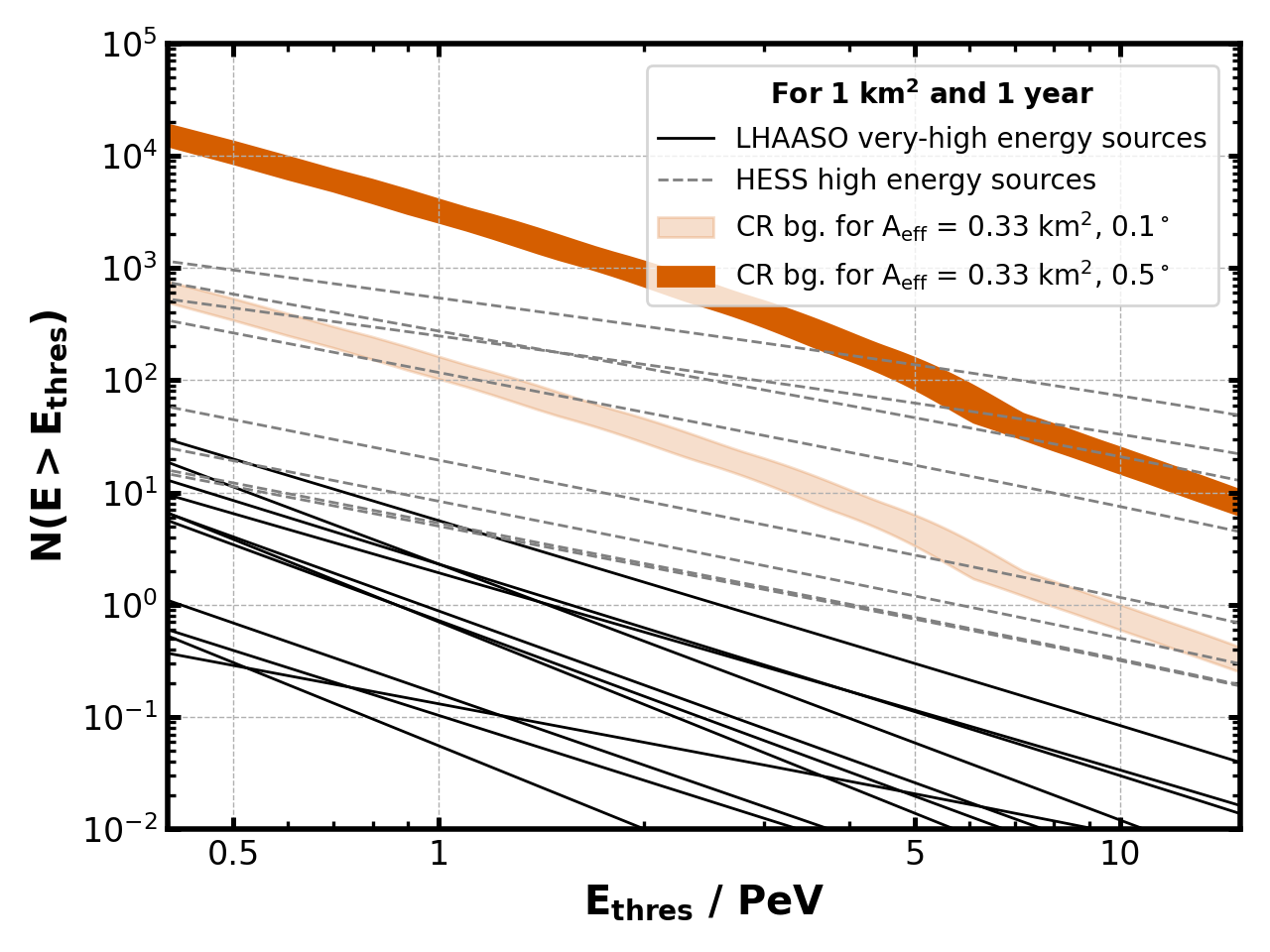}
    \caption{Number of expected showers above a given energy $E$ as function of threshold detection energy for an area corresponding to the entire SKA-Low core, which is a simplified assumption. Shown are expectations for various $\gamma$-ray sources, as well as the cosmic-ray background. Figure from \cite{ICRC_2025_Felix}.}
    \label{fig:fluxes}
\end{figure}

SKA-Low offers significant advantages in angular resolution and energy-reconstruction accuracy. These features will allow deep, high-resolution spectral and morphological studies of known TeV sources to robustly confirm their extension into the PeV regime and potentially lead to the discovery of new PeVatron classes. However, a key challenge remains its yet unknown rejection power against the hadronic cosmic-ray background compared to established techniques as used by, for instance, LHAASO. Figure \ref{fig:fluxes} illustrates the signal expected from specific sources, compared to the expected background rates from hadronic cosmic rays. We note that extrapolation to the highest energies comes with large uncertainties and that the assumed effective area is a coarse estimate at this moment that needs to be refined further. 

The SKA-Low observations combined with multi-wavelength observations will be essential to constrain emission mechanisms and firmly identify PeVatron candidates through morphological and spectral modeling.

Given the already dense instrumentation foreseen in the inner core for the SKA-Low staged delivery of AA*, we anticipate to be able to reach all of these science goals with AA*, given enough time allocation. Only a fraction of the antennas will be needed (and available) for the raw trigger read-out and the time allocation will depend on the fraction of commensal observations, which is currently not well-specified.  We do note that the denser the array, the easier it will be to run observations of sub-arrays in parallel, which is why it will also be beneficial for the science case of $\gamma$-ray detection to have AA4 available.




\section{Radio Detection of Air Showers with SKA}

\begin{figure}
    \centering
    \includegraphics[width=0.49\linewidth]{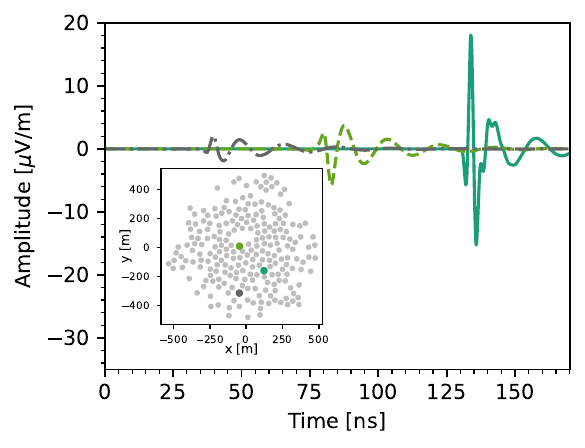}
    \includegraphics[width=0.49\linewidth]{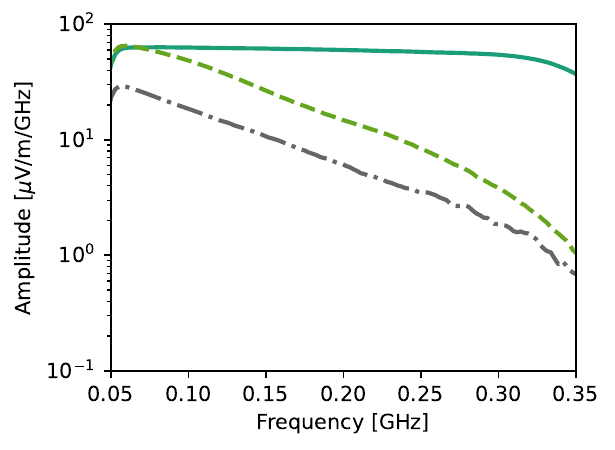}
    \caption{Illustration of noiseless air shower signals to be measured with a single SKA-Low antenna. Left: Time-domain signals as expected in one antenna (location in the antenna array marked in the inset; the shower core is at (0, 0)). Right: Corresponding frequency spectrum. Different frequency spectra correspond to different distances from the shower axis. Shown is the $E_\theta$ electric field component for simplicity, pulse shapes per antenna will be similar for other projections. Note that the bandpass of SKA-Low has been approximated and the final pulse shape will change for extremely broad-band pulses.}
    \label{fig:pulses}
\end{figure}

Radio detection of cosmic rays is a mature concept, being routinely used at various experiments, such as the Pierre Auger Observatory. It relies on measuring the nanosecond-scale radio pulses emitted from the charged particles in the air shower that follow the interaction of a cosmic ray in the atmosphere. The amplitude of the radio emission is essentially governed by the electrons and positrons in the shower, which means that radio emission is expected from all showers, irrespective of the primary particle. The field amplitude scales with the energy of the primary particle. Typical pulses as expected from air showers in the SKA-Low bandwidth are shown in Figure \ref{fig:pulses}.

\subsection{Detection concept}

For SKA-Low, it is first planned to focus on air showers induced by hadronic cosmic rays, which is discussed in more detail in \citep{SKA_Book_Cosmic_Rays}. For this, a particle detector array is envisioned within the core of SKA-Low \citep{SKA_Book_CR_Overview}. The presence of particles will ensure that the raw voltage waveforms of SKA-Low that have been filled in buffers will only be read-out, if an air shower occurred. A well-tuned triggering algorithm that combines a simple pulse finding with a particle coincidence will essentially reduce the false-positives to zero, as it has been shown with LOFAR \citep{Schellart:2013bba}. The typical observation threshold that has been achieved with LOFAR is about 50 PeV, thus, well-above the energy regime interesting for $\gamma$-ray observations. The reason being that the pulse amplitude does not significantly exceed the Galactic noise in the band of 30 - 80 MHz for these shower energies, making triggering and reconstruction challenging. 

This chapter deals with a potential enhancement to this baseline performance, a dedicated triggering pipeline for $\gamma$-rays. Given the opportunities in being one of the few experiments and the first ever radio telescope to observe a $\gamma$-ray flux at PeV energies, it is worth to entertain the idea of what might be needed. 
The SWG High Energy Cosmic Particles is currently also exploring options in how far it will be feasible to combine the particle trigger, with a more complex radio-based trigger to lower the energy threshold of all air shower observations. This will benefit both the $\gamma$-ray case, as well as broaden the energy range in which SKA-Low can contribute to hadronic (air shower) physics (cf.~\cite{SKA_Book_CR_Overview}). 

A notable aspect of $\gamma$-ray observation compared to hadronic air showers is that one can choose a source of interest and therefore the arrival direction is fixed apriori. We expect SKA, owing to its nature of a radio interferometry and the thus required timing stability, to have an air shower angular resolution of at least \ang{0.1} \citep{Corstanje:2014waa}, which allows to precisely target a source. This means that one can pre-select an arrival direction in the trigger, which will immensely reduce the background of hadronic cosmic rays as already alluded to in Figure \ref{fig:fluxes} and reduce the trigger complexity. 

\subsection{Energy threshold}

\begin{figure}
    \centering
    \includegraphics[trim={0 0 0 0},clip,width=0.49\linewidth]{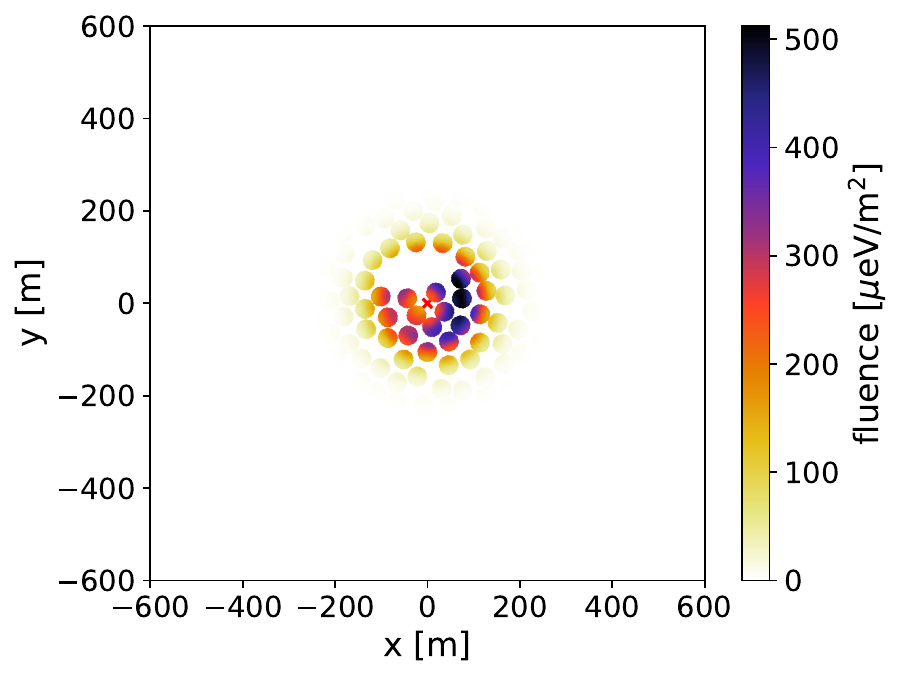}
     \includegraphics[trim={0 0 0 0},clip,width=0.49\linewidth]{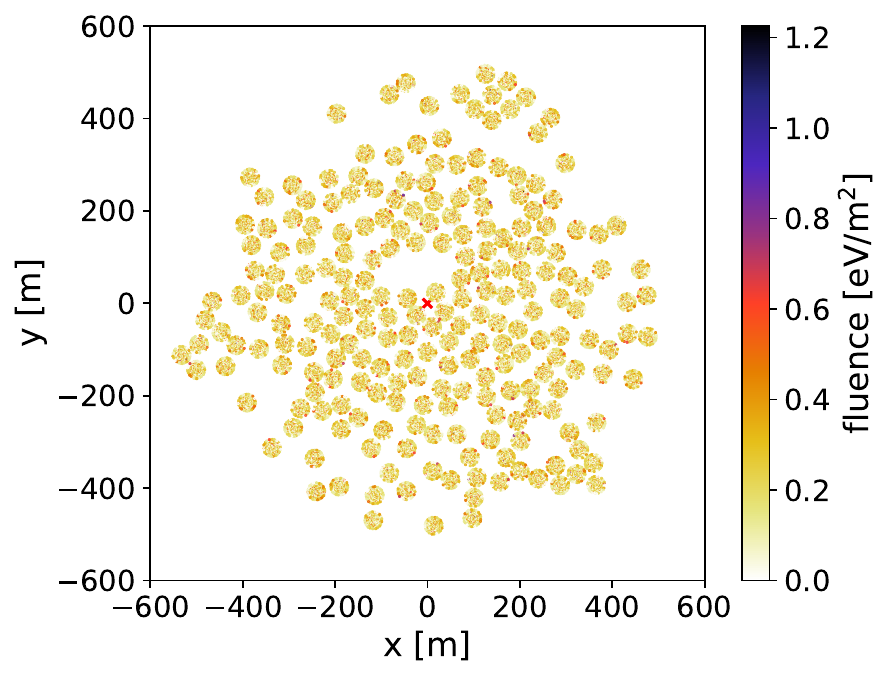}
    \includegraphics[trim={0 0 0 0},clip,width=0.49\linewidth]{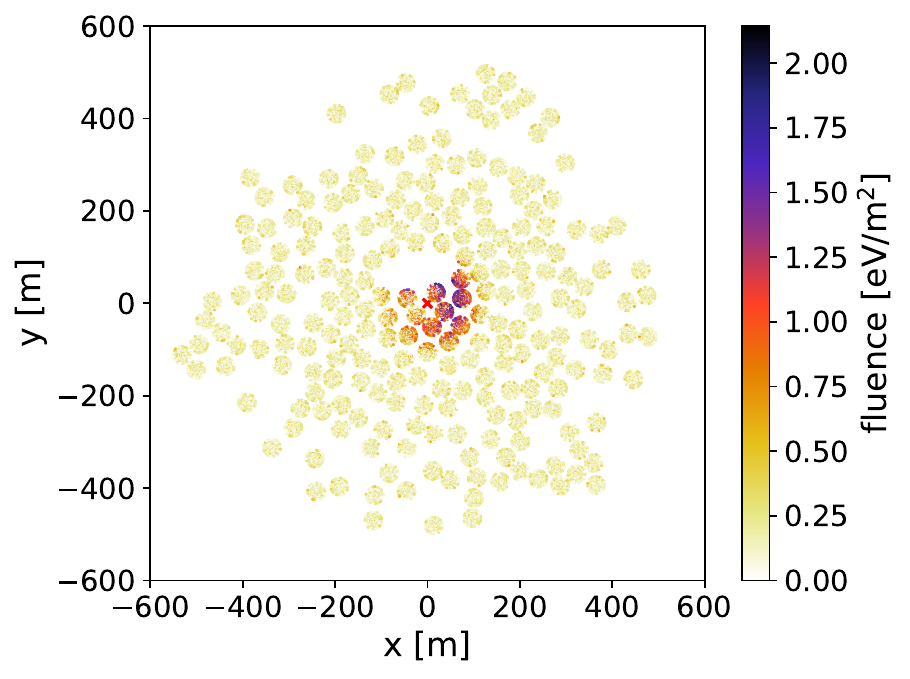}
     \includegraphics[trim={0 0 0 0},clip,width=0.49\linewidth]{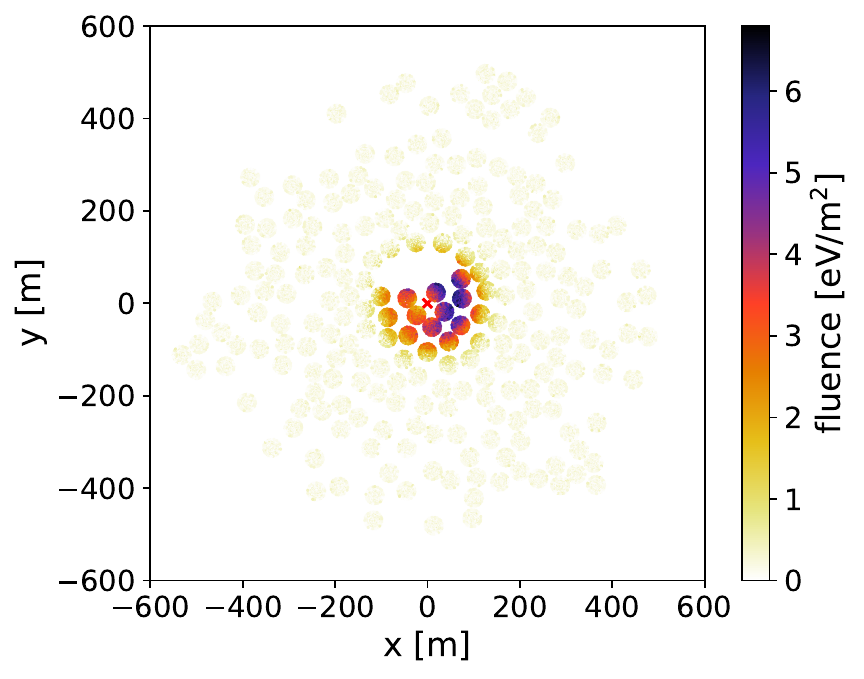}
    \caption{Simulated footprints from air showers initiated by a $\gamma$-ray at 15 degrees zenith angle as they would be detected with SKA-Low. Top row, left: Shower energy 0.5 PeV, no noise; right: Same shower with instrumental and Galactic noise added. Bottom row: the same shower (arrival direction), but at 25 PeV (left) and 50 PeV (right), both including noise. }
    \label{fig:footprints}
\end{figure}

Figure \ref{fig:footprints} illustrates the challenges in lowering the energy threshold of the detection. While an air shower induced by a $\gamma$-ray is clearly visible in simulations already at 0.5 PeV, it essentially becomes undetectable when instrumental noise and Galactic diffuse emission is added\footnote{The precise energy threshold will be subject to the in-situ system performance.}. Only at around 25 PeV would a simple pulse finding algorithm be able to retrieve pulses to ensure an appropriate trigger. While triggering on PeV showers using particles is in principle not a problem, the hadronic cosmic ray flux follows a power-law that is a steeply falling function of energy $\sim E^{-3}$. This means that lowering the required energy threshold in a particle-only trigger immediately swamps the read-out of SKA-Low to multiple triggers per second, if no further down-selection happens. 

As is well known in radio interferometry, beamforming enables us to retrieve signal from the noise. The beamforming that is used for air showers is, however, subtly different from the classical time-integrated far-field beamforming. 
First of all, the shower takes place in the atmosphere at heights of $\mathcal{O}(10)$ km above ground. This means that at a baseline of \SI{1}{km}, the angle at which different antennas see the shower is significantly different by up to several degrees. 
Furthermore, an air shower is not a point source, but the emission is extended, making the observed signals follow a hyperbolic wavefront \citep{Corstanje:2014waa}. Also, the emission is forward beamed, which means that only a fraction of the antennas receive a signal (cf. Figure \ref{fig:footprints}), which means that simply adding more antennas can lead to a worse beamformed result. 

Several approaches have recently explored the practical applicability of beamforming for air showers and the mathematical peculiarities \citep{Scholten:2024upn,Schluter:2021egm,Schoorlemmer:2020low}, and the reader is referred to these publications for a in-depth description. For here, it suffices to say that we use a method referred to as \emph{RIT} \citep{Schoorlemmer:2020low,Schluter:2021egm}, where the beamformed signal at a given location in the atmosphere $\vec{j}$ is defined as
\begin{equation}
    B_j(t) = \sum_i^\text{ant} S_i (t-\Delta_{i,j}),
    \label{eq:beam}
\end{equation}
where  $S_i$ are the signals in each antenna and the $\Delta_{i,j}$ is the time-shift according to the location, taking into account varying atmospheric density which affects the signal travel time. If the location $\vec{j}$ is close to the main emission region, the signals will be adding up coherently and thus provide a large pulse as function of time. In contrast to imaging, in particular, no time-integration is performed. 
Work is ongoing to ultimately perform a 3-dimensional reconstruction of an air shower \citep{Straub:2025lsd,Watanabe:2025njo} offline, in which detailed characteristics of the shower can be extracted. However, for an online trigger as it would be implemented for $\gamma$-ray observations, such a complex reconstruction is too time consuming, which is why this chapter discusses simplified ideas suitable for an online system.

\begin{figure}
    \centering
    \includegraphics[width=0.5\linewidth]{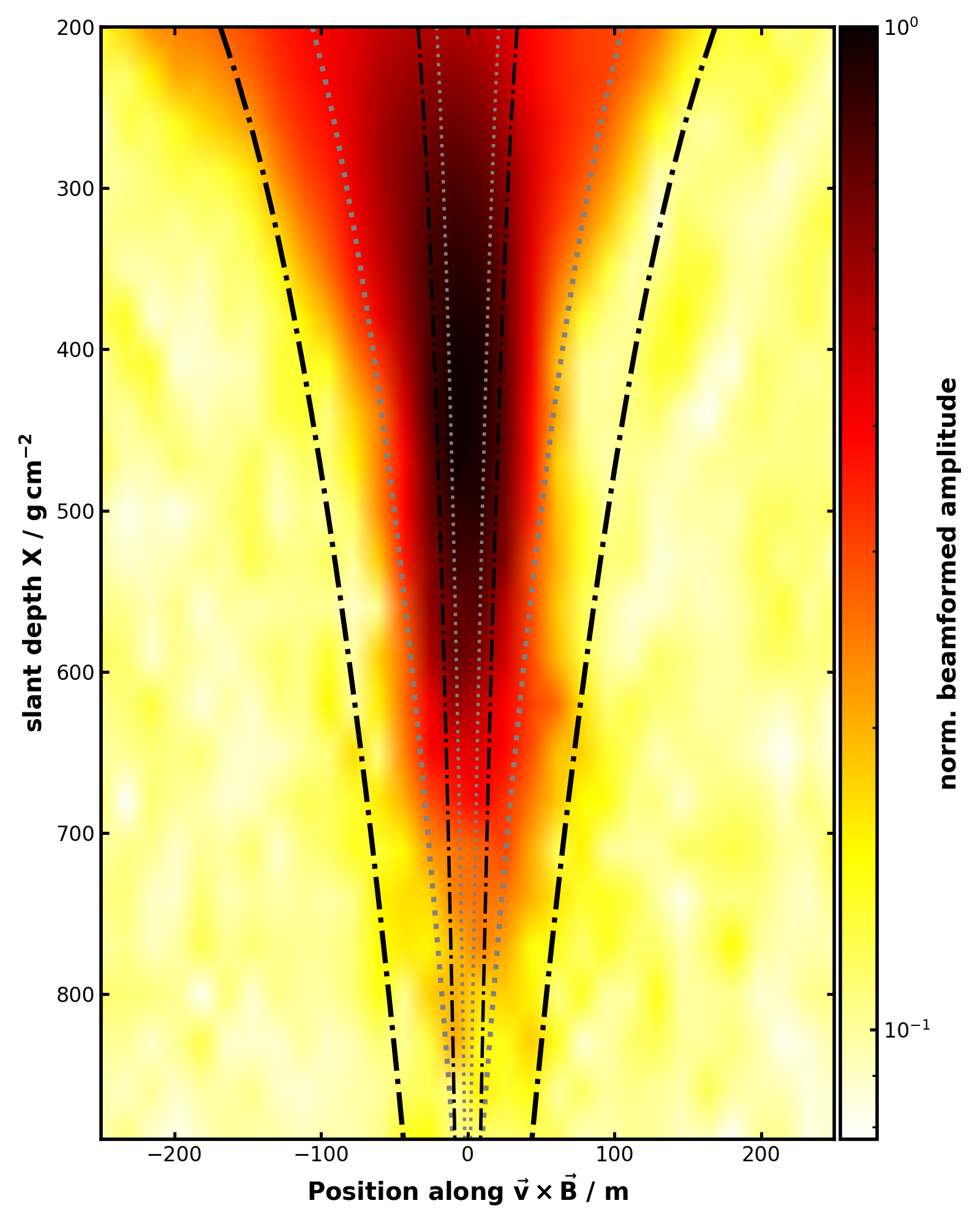}
    \caption{Beamformed signal amplitude (see Equation \ref{eq:beam}) as function of atmospheric depth and slice along the shower axis, defined by shower axis ($\vec{v}$) and geomagnetic field ($\vec{B}$) for one example air shower as detected with SKA-Low. The shower development from the top of the atmosphere ($X = 0$) to the observer height is imaged in this simplified beamforming. }
    \label{fig:rit}
\end{figure}

\begin{figure}
    \centering
    \includegraphics[width=0.49\linewidth]{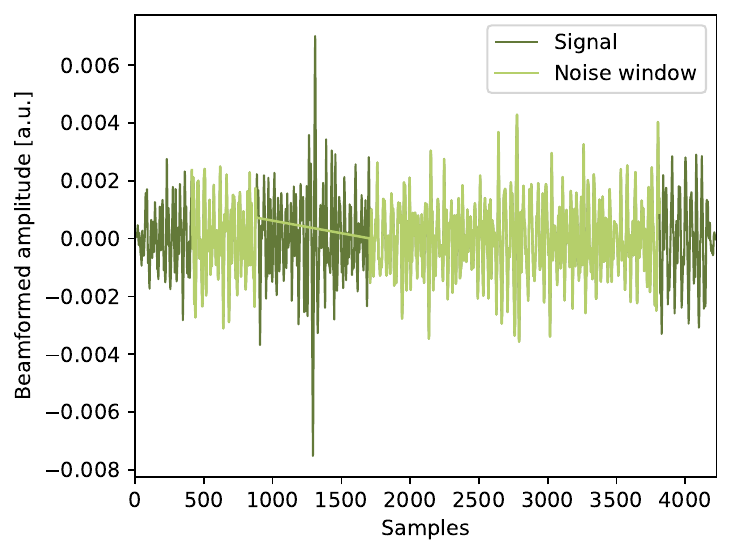}
    \includegraphics[width=0.49\linewidth]{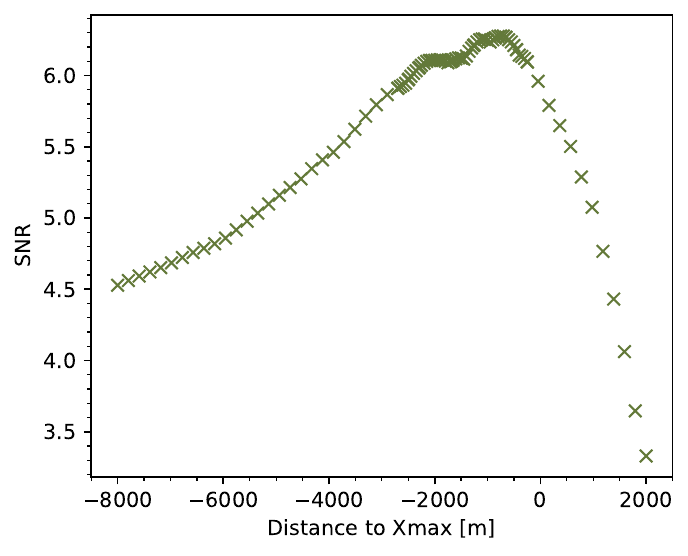}
    \caption{Left: Best beamformed trace (dark green) from scan along the shower axis for a $300$\,TeV gamma ray shower with a zenith angle of $15\,^\circ$ (simulation). The noise parts used for noise RMS calculation are marked in light green, excluding both the pulse and edge effects. Right: Obtained Beamforming signal-to-noise ratio (SNR) as function of distance to shower maximum along the shower axis for an example gamma ray simulation. Each point in this panel corresponds to the SNR of one waveform like shown in the left. Close to $X_{\text{max}}$ the sampling step size has been reduced in this figure.}
    \label{fig:example_best_beamformed_trace_scan}
\end{figure}

Figure \ref{fig:rit} shows a two-dimensional slice of a beamformed air shower following the method implemented by \cite{Schluter:2021egm}. Each point in this map corresponds to the maximum amplitude of one beamformed waveform as shown in Figure \ref{fig:example_best_beamformed_trace_scan}. It should be noted that the shower has been transformed to a coordinate system that allows a general treatment across different experiment and air showers. This coordinate system is defined by the shower axis $\vec{v}$ and the geomagnetic field $\vec{B}$. Also, the shower is not shown in its geometric height, but in column density of the atmosphere traversed, where \SI{0}{g\ cm^{-2}} is the top of the atmosphere. A typical quantity in air shower physics refers to the \emph{shower maximum}, $X_{\text{max}}$, also given in these units, that describes the height at which the shower reaches its maximal extent in numbers of particles. 

While it is for air shower physics in principle interesting to see the lateral extent of the shower (x-axis in Figure \ref{fig:rit}), this is not needed for a trigger. Therefore one can restrict oneself to a scan along the shower axis (y-axis), which is shown on the right in Figure \ref{fig:example_best_beamformed_trace_scan}. The Figure shows that, if one knows the shower axis (arrival direction and where the core of the shower arrived on ground) precisely and can perform such a scan along the atmospheric depth, given here as distance to the shower maximum. Very close to the shower maximum, also the signal-to-noise-ratio (SNR) of the beamformed pulse reaches its maximum. We define SNR as maximum amplitude defined by the RMS of the noise (see Figure \ref{fig:example_best_beamformed_trace_scan}). Given past experience with triggering, it is safe to assume that an SNR of about 5 is sufficient to trigger on, but field tests will have to be conducted.  

\begin{figure}
    \centering
    \includegraphics[width=0.49\linewidth]{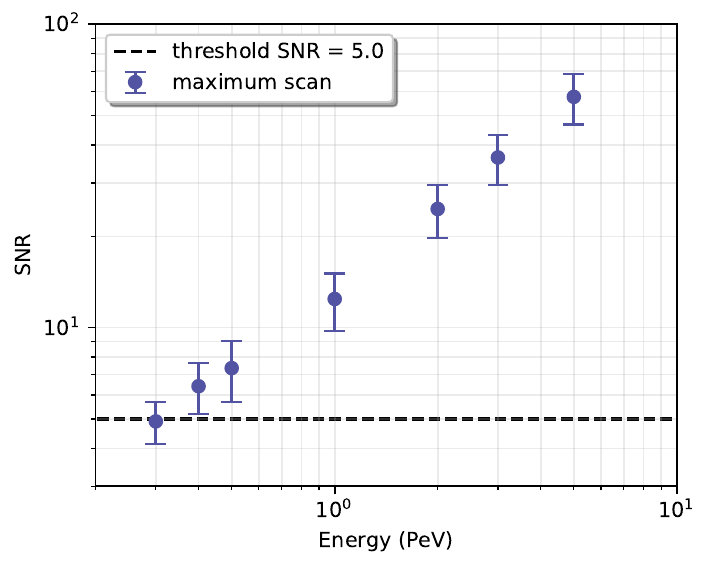}
        \includegraphics[width=0.49\linewidth]{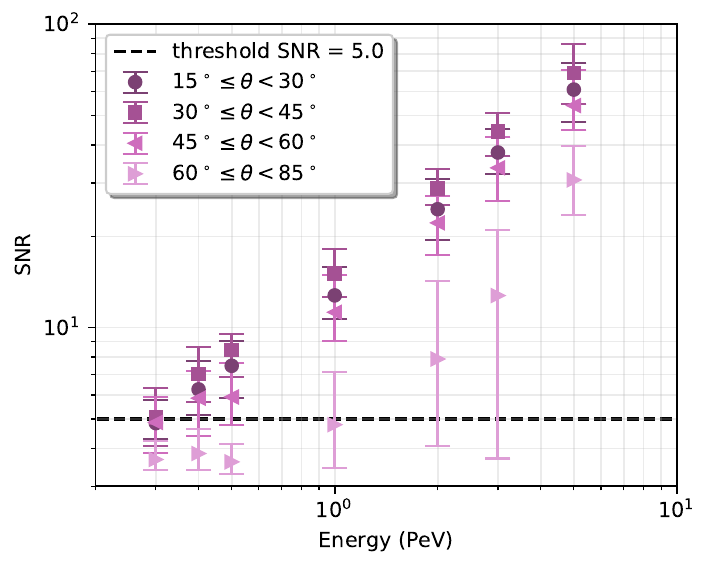}
    \caption{Signal-to-noise ratio (SNR) of the best beamformed signal as function of the simulated gamma ray energy for a zenith angle of \ang{15} (left) and several different zenith angle bins (right). The zenith angle is defined as \ang{0} pointing upwards. The shower core lies within 400 meters of the SKA array core (quality cut). Antennas within 200 meters from the shower axis are used for beamforming. The shower axis is assumed to be known.}
    \label{fig:SNR_vs_energy_zen}
\end{figure}

Since the electric field strength received per antenna essentially scales with the shower energy, it is easier to detect higher energy showers and the minimum obtainable SNR determined the energy threshold of the experiment. Figure \ref{fig:SNR_vs_energy_zen} shows the SNR as a function of energy for different zenith angles. When knowing the shower axis, a detection of a \SI{1}{PeV} shower seems feasible, possibly pushing towards lower energies. The performance is similar for all sources with an elevation of at least \ang{30}, which means that some sources will be more easy to observe with SKA in $\gamma$-rays than others. 

\begin{figure}
    \centering
    \includegraphics[width=0.6\linewidth]{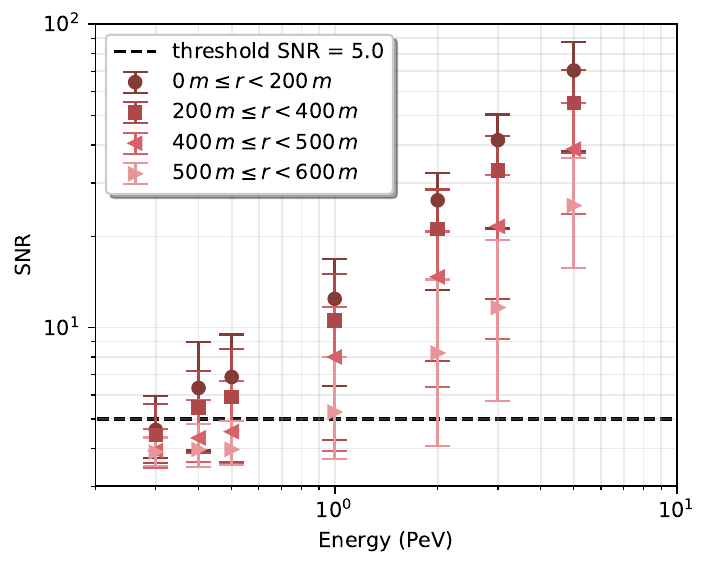}
    \caption{SNR of the best beamformed signal as a function of the simulated gamma ray energy for different distances between shower core and SKA array core, see legend. Antennas within 200 meters from the shower axis are used for beamforming. The shower axis is assumed to be known.}
    \label{fig:SNR_vs_energy_core_dist_bins}
\end{figure}

In the previous examples, the shower axis could lie almost everywhere inside the antenna array with the only constraint that it lies within \SI{400}{m} from the SKA-Low array core. Figure \ref{fig:SNR_vs_energy_core_dist_bins} shows in detail how the beamforming SNR behaves for different distances of the shower axis to center of the antenna array i.e.\ how contained the shower must be in order to detect a signal using beamforming. At least out to \SI{500}{m} the method continues to be very efficient, with still some efficiency at higher energies further out. This is the first step towards establishing an effective area calculation, which will allow us to better quantify the expected number of events (see Figure \ref{fig:fluxes}). 

All of these figures assume that the shower axis is known a-priori. In a realistic setting, however, this will not be the case since air shower randomly fall on ground. Therefore, one also has to scan for that and establish a good algorithm, which is currently being developed. Only after the conclusion of these methodological studies, one can advance to drawing firm conclusions about the feasibilty. 

\subsection{Gamma-Hadron Separation}

As alluded to in Figure \ref{fig:fluxes}, a detection of a $\gamma$-ray source has to be done against the background of hadronic cosmic rays. The superior angular resolution of SKA, will allow to suppress a significant fraction, however, at PeV energies most models still expect more hadronic cosmic rays than gamma rays. 

\begin{figure}
    \centering
    \includegraphics[width=0.6\linewidth]{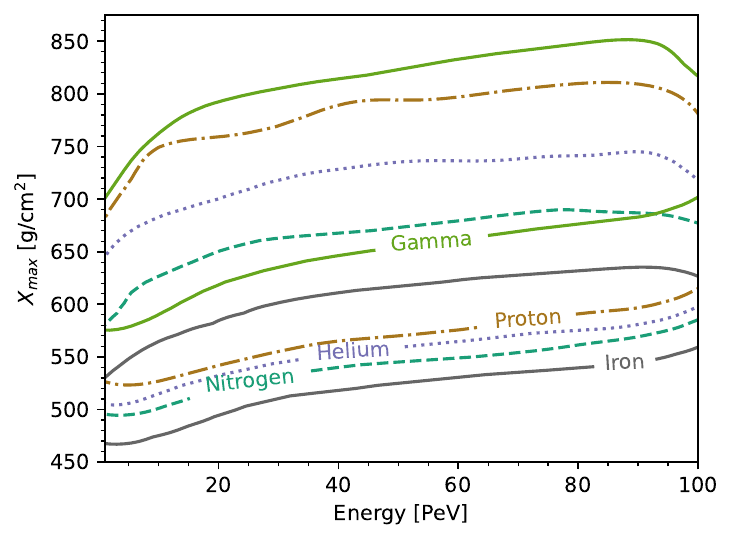}
    \caption{Height of shower maximum as function of energy for simulated air showers of different primary particle. Shown are the 90\,\% contours for different primaries. These results are based on CORSIKA simulations with CONEX using approx. 2500 showers per primary.} 
    \label{fig:xmax_vs_energy}
\end{figure}

It is known that based on $X_{\text{max}}$ alone, some fraction of hadronic showers can be suppressed (see Figure \ref{fig:xmax_vs_energy}), but in particular protons are hard to distinguish from $\gamma$-rays. In imaging air-Cherenkov telescopes like H.E.S.S. the smoothness of the Cherenkov light is used to suppress background. In particle detectors, like LHAASO, the muon content of the shower aids the background suppression. In radio detection of air showers, no one has studied the capabilities in detail. Since radio emission is known to be sensitive to the shower profile, there may be additional leverage that can be explored with SKA-Low, which is expected to show superior performance in the profile reconstruction \citep{Corstanje:2025wbc}. 

\section{Open Questions towards Gamma-Ray Detection}
\label{sec:future}

Compared to $\gamma$-ray detection, the strategy of detecting hadronic cosmic rays with SKA-Low is much more mature. This means that a large number of questions are still open and subject to future work. We list a number of them here. 

\paragraph{How will the trigger work in practice:} It is clear that $\gamma$-ray detection will mainly be driven by a trigger based on the radio signal itself. Such a self-trigger has never been implemented at LOFAR, because a particle-based trigger was much more robust. For $\gamma$-rays, the energy threshold has to be pushed lower, which means that a particle-based trigger is less efficient and at least a hybrid-trigger using both radio and the particles will be needed. How a hybrid trigger will or could be designed is still very much under discussion. 

Work at the OVRO-LWA has shown that a self-trigger can be efficient \citep{Monroe:2019zkp}, but that each trigger has to be tuned to the instrument and the specific site. Therefore an SNR of 5 is used as proxy in this study, but it will require in-field tests to see whether this is feasible and will not lead to too many false positive triggers. This threshold will also determine the $\gamma$-ray energy that can be targeted, in particular methods of beam-forming are still being developed for the case of an unknown shower geometry. 

\paragraph{How efficient will the gamma-hadron separation work:} Up to now, no one has worked on algorithms that can provide a separation of $\gamma$-ray induced showers using the radio signal of an air shower. Dedicated arrays like H.E.S.S. or LHAASO, built on years of experience and method development, exploiting features in the Cherenkov light and the particle content of the shower. Initial work suggests that radio and its ability to reconstruct the full shower profile will be a powerful tool, but it is too early to say whether this can be successful. 

In parallel, it is being explored whether the particle array can be enhance to aid in the gamma-hadron separation. Also here a number of questions remain open and will have to be addressed in future work. 

\paragraph{Is there a significant flux of $\gamma$-rays at $> 5$ PeV energies:} LHAASO has shown that the highest energy $\gamma$-rays were detected at a couple of PeV, which may just be out of reach towards the low energies at SKA. It is argued that the LHAASO-sources have a cut-off or at least a spectral turn-over at the highest energies, which will make a detection harder still. For an experiment like SKA-Low that aims at discovery, the discovery also relies on nature providing a signal, while we note that also upper-limits can provide useful constraints.

\section{Conclusions}

To summarize, the science case of providing the first observation of a $\gamma$-ray shower through its radio emission and aiding in the understanding of the most powerful particle accelerators in our Galaxy is strong. With a detection, SKA-Low would enter the regime of dedicated detectors such as LHAASO, H.E.S.S. or the upcoming CTAO. 
From a technical perspective, however, many questions remain open and work has just begun to flesh out triggering approaches, particle identification, and data flow. At this point, the work towards $\gamma$-ray detection is highly synergetic with the work towards the detection of hadronic cosmic rays. The challenges posed by $\gamma$-ray detection will act as catalyzing agent to develop new methods to lower the detection threshold and improve particle identification.

\section*{Acknowledgments}

The authors build on countless efforts to enable air shower observations using radio emission and are indebted to the community. Concretely, we acknowledge the following support: 
SBo, AN, and KT acknowledge the Verbundforschung of the German Ministry for Research, Technology and Space (BMFTR). 
PL and KW are supported by the Deutsche Forschungsgemeinschaft (DFG, German Research Foundation) – Projektnummer 531213488.
BH, CS, and PT are supported by ERC Grant Agreement No.~101041097.
KM acknowledges funding from the Netherlands Research School for Astronomy (NOVA) and Dutch Research Council (NWO) project OCENW.XS25.1.237. ST acknowledges funding from the Khalifa University RIG-S-2023-070 grant. This research is supported by the Flemish Foundation for Scientific Research (FWO-AL991 and FWO-OZR4291).
The authors gratefully acknowledge the computing time provided on the high-performance computer HoreKa by the National High-Performance Computing Center at KIT (NHR@KIT). This center is jointly supported by the Federal Ministry of Education and Research and the Ministry of Science, Research and the Arts of Baden-Württemberg, as part of the National High-Performance Computing (NHR) joint funding program. HoreKa is partly funded by the German Research Foundation.

\bibliographystyle{abbrvnat-maxbibnames4}
\bibliography{chapter} 

\begin{thebibliography}{42}
\providecommand{\natexlab}[1]{#1}
\providecommand{\url}[1]{\texttt{#1}}
\expandafter\ifx\csname urlstyle\endcsname\relax
  \providecommand{\doi}[1]{doi: #1}\else
  \providecommand{\doi}{doi: \begingroup \urlstyle{rm}\Url}\fi

\bibitem[Aab et~al.(2016)]{PierreAuger:2015hbf}
A.~Aab et~al.
\newblock \emph{Phys. Rev. D}, 93\penalty0 (12):\penalty0 122005, 2016.
\newblock \doi{10.1103/PhysRevD.93.122005}.

\bibitem[Abdalla et~al.(2018)]{HESS:2018pbp}
H.~Abdalla et~al.
\newblock \emph{Astron. Astrophys.}, 612:\penalty0 A1, 2018.
\newblock \doi{10.1051/0004-6361/201732098}.

\bibitem[Abeysekara et~al.(2017)]{Abeysekara:2017hyn}
A.~U. Abeysekara et~al.
\newblock \emph{Astrophys. J.}, 843\penalty0 (1):\penalty0 40, 2017.
\newblock \doi{10.3847/1538-4357/aa7556}.

\bibitem[Aharonian et~al.(2024)]{HESS:2024rlh}
F.~Aharonian et~al.
\newblock \emph{Science}, 383\penalty0 (6681):\penalty0 adi2048, 2024.
\newblock \doi{10.1126/science.adi2048}.

\bibitem[Aharonian et~al.(2025)]{HESS:2025gpm}
F.~Aharonian et~al.
\newblock \emph{Astron. Astrophys.}, 695:\penalty0 A261, 2025.
\newblock \doi{10.1051/0004-6361/202452723}.

\bibitem[Alfaro et~al.(2024)]{Alfaro:2024cjd}
R.~Alfaro et~al.
\newblock \emph{Nature}, 634:\penalty0 557--560, 2024.
\newblock \doi{10.1038/s41586-024-07995-9}.

\bibitem[Amenomori et~al.(2023)]{Amenomori2023}
M.~Amenomori et~al.
\newblock \emph{Astrophysical Journal}, 954\penalty0 (2):\penalty0 200, Sept.
  2023.
\newblock \doi{10.3847/1538-4357/acebce}.

\bibitem[Apel et~al.(2016)]{Tunka-Rex:2016nto}
W.~D. Apel et~al.
\newblock \emph{Phys. Lett. B}, 763:\penalty0 179--185, 2016.
\newblock \doi{10.1016/j.physletb.2016.10.031}.

\bibitem[Braun et~al.(2019)Braun, Bonaldi, Bourke, Keane, and
  Wagg]{braun2019anticipatedperformancesquarekilometre}
R.~Braun et al.
\newblock {Anticipated Performance of the Square Kilometre Array -- Phase 1
  (SKA1)}, 2019.
\newblock URL \url{https://arxiv.org/abs/1912.12699}.

\bibitem[Buitink et~al.(2026)Buitink, author2, author3, author4, and
  author5]{SKA_Book_Hadronic}
S.~Buitink et al.
\newblock In \emph{Advancing Astrophysics with the SKA -- II (AASKAII)}. 2026.
\newblock arXiv search: Report number AASKAII/Buitink01.

\bibitem[Cao et~al.(2021{\natexlab{a}})]{LHAASO:2021Nat}
Z.~Cao et~al.
\newblock \emph{Nature}, 594\penalty0 (7861):\penalty0 33--36, June
  2021{\natexlab{a}}.
\newblock \doi{10.1038/s41586-021-03498-z}.

\bibitem[Cao et~al.(2021{\natexlab{b}})]{LHAASO:2021cbz}
Z.~Cao et~al.
\newblock \emph{Science}, 373\penalty0 (6553):\penalty0 425--430,
  2021{\natexlab{b}}.
\newblock \doi{10.1126/science.abg5137}.

\bibitem[Cao et~al.(2021{\natexlab{c}})]{LHAASO:2021plb}
Z.~Cao et~al.
\newblock \emph{Astrophys. J. Lett.}, 917\penalty0 (1):\penalty0 L4,
  2021{\natexlab{c}}.
\newblock \doi{10.3847/2041-8213/ac0fd5}.

\bibitem[Cao et~al.(2024{\natexlab{a}})]{LHAASO:2023rpg}
Z.~Cao et~al.
\newblock \emph{Astrophys. J. Suppl.}, 271\penalty0 (1):\penalty0 25,
  2024{\natexlab{a}}.
\newblock \doi{10.3847/1538-4365/acfd29}.

\bibitem[Cao et~al.(2024{\natexlab{b}})]{LHAASO:2023uhj}
Z.~Cao et~al.
\newblock \emph{Sci. Bull.}, 69\penalty0 (4):\penalty0 449--457,
  2024{\natexlab{b}}.
\newblock \doi{10.1016/j.scib.2023.12.040}.

\bibitem[Cao et~al.(2024{\natexlab{c}})]{LHAASO:2024fym}
Z.~Cao et~al.
\newblock \emph{Sci. Bull.}, 69:\penalty0 2833--2841, 2024{\natexlab{c}}.
\newblock \doi{10.1016/j.scib.2024.07.017}.

\bibitem[Cao et~al.(2025)]{LHAASO:2024lls}
Z.~Cao et~al.
\newblock \emph{Sci. China Phys. Mech. Astron.}, 68\penalty0 (7):\penalty0
  279502, 2025.
\newblock \doi{10.1007/s11433-024-2477-9}.

\bibitem[Cardillo and Giuliani(2023)]{Cardillo:2023hbb}
M.~Cardillo and A.~Giuliani.
\newblock \emph{Appl. Sciences}, 13\penalty0 (11):\penalty0 6433, 2023.
\newblock \doi{10.3390/app13116433}.

\bibitem[Chang et~al.(2022)Chang, Zhang, and Zhou]{Chang:2022fvj}
Z.~Chang, X.~Zhang, and J.-Z. Zhou.
\newblock \emph{Mon. Not. Roy. Astron. Soc.}, 516\penalty0 (4):\penalty0
  4916--4921, 2022.
\newblock \doi{10.1093/mnras/stac2553}.

\bibitem[Corstanje et~al.(2015)Corstanje, Schellart, Nelles,
  et~al.]{Corstanje:2014waa}
A.~Corstanje, P.~Schellart, A.~Nelles, et~al.
\newblock \emph{Astropart. Phys.}, 61:\penalty0 22--31, 2015.
\newblock \doi{10.1016/j.astropartphys.2014.06.001}.

\bibitem[Corstanje et~al.(2026)Corstanje, author2, author3, author4, and
  author5]{SKA_Book_Cosmic_Rays}
A.~Corstanje et al.
\newblock In \emph{Advancing Astrophysics with the SKA -- II (AASKAII)}. 2026.
\newblock arXiv search: Report number AASKAII/Corstanje01.

\bibitem[Corstanje et~al.(2025)]{Corstanje:2025wbc}
A.~Corstanje et~al.
\newblock \emph{Phys. Rev. D}, 112\penalty0 (2):\penalty0 023017, 2025.
\newblock \doi{10.1103/l8mt-994v}.

\bibitem[de~O{\~n}a~Wilhelmi et~al.(2024)de~O{\~n}a~Wilhelmi, L{\'o}pez-Coto,
  Aharonian, Amato, Cao, Gabici, and Hinton]{deOnaWilhelmi:2024mul}
E.~de~O{\~n}a~Wilhelmi et al.
\newblock \emph{Nature Astron.}, 8\penalty0 (4):\penalty0 425--431, 2024.
\newblock \doi{10.1038/s41550-024-02224-9}.

\bibitem[De~Sarkar and Gupta(2022)]{DeSarkar:2022bnl}
A.~De~Sarkar and N.~Gupta.
\newblock \emph{Astrophys. J.}, 934\penalty0 (2):\penalty0 118, 2022.
\newblock \doi{10.3847/1538-4357/ac6ce5}.

\bibitem[Dom{\'\i}nguez et~al.(2013)Dom{\'\i}nguez, Finke, Prada, Primack,
  Kitaura, Siana, and Paneque]{Dominguez:2013lfa}
A.~Dom{\'\i}nguez et al.
\newblock \emph{Astrophys. J.}, 770:\penalty0 77, 2013.
\newblock \doi{10.1088/0004-637X/770/1/77}.

\bibitem[Hu et~al.(2025)Hu, Fang, and Collaboration]{Hu:2025SD}
S.~Hu, K.~Fang, and L.~Collaboration.
\newblock \emph{PoS}, ICRC2025:\penalty0 681, 2025.
\newblock \doi{10.22323/1.501.0681}.

\bibitem[{LHAASO Collaboration}(2024)]{LHAASO:2024psv}
{LHAASO Collaboration}.
\newblock \emph{arXiv e-prints}, art. arXiv:2410.08988, Oct. 2024.
\newblock \doi{10.48550/arXiv.2410.08988}.

\bibitem[Li et~al.(2025)]{Li:2025NF}
H.~Li et~al.
\newblock \emph{PoS}, ICRC2025:\penalty0 733, 2025.
\newblock \doi{10.22323/1.501.0733}.

\bibitem[Mitchell and Celli(2024)]{Mitchell:2024oue}
A.~M.~W. Mitchell and S.~Celli.
\newblock \emph{JHEAp}, 44:\penalty0 340--355, 2024.
\newblock \doi{10.1016/j.jheap.2024.10.011}.

\bibitem[Monroe et~al.(2020)]{Monroe:2019zkp}
R.~Monroe et~al.
\newblock \emph{Nucl. Instrum. Meth. A}, 953:\penalty0 163086, 2020.
\newblock \doi{10.1016/j.nima.2019.163086}.

\bibitem[Mulrey et~al.(2026)Mulrey, author2, author3, author4, and
  author5]{SKA_Book_CR_Overview}
K.~Mulrey et al.
\newblock In \emph{Advancing Astrophysics with the SKA -- II (AASKAII)}. 2026.
\newblock arXiv search: Report number AASKAII/Huege01.

\bibitem[Peron et~al.(2024)Peron, Morlino, Gabici, Amato, Purushothaman, and
  Brusa]{Peron:2024lfo}
G.~Peron et al.
\newblock \emph{Astrophys. J. Lett.}, 972\penalty0 (2):\penalty0 L22, 2024.
\newblock \doi{10.3847/2041-8213/ad7024}.

\bibitem[Sabri et~al.(2025)Sabri, Gallant, Devin, and Feijen]{Sabri:2025qxd}
K.~Sabri, Y.~Gallant, J.~Devin, and K.~Feijen.
\newblock \emph{PoS}, ICRC2025:\penalty0 828, 2025.
\newblock \doi{10.22323/1.501.0828}.

\bibitem[Schellart et~al.(2013)Schellart, Nelles, Buitink,
  et~al.]{Schellart:2013bba}
P.~Schellart, A.~Nelles, S.~Buitink, et~al.
\newblock \emph{Astron. Astrophys.}, 560:\penalty0 A98, 2013.
\newblock \doi{10.1051/0004-6361/201322683}.

\bibitem[Schl{\"u}ter and Huege(2021)]{Schluter:2021egm}
F.~Schl{\"u}ter and T.~Huege.
\newblock \emph{JINST}, 16\penalty0 (07):\penalty0 P07048, 2021.
\newblock \doi{10.1088/1748-0221/16/07/P07048}.

\bibitem[Schlüter et~al.(2025)Schlüter, Laub, Nelles, Bouma, Bray,
  et~al.]{ICRC_2025_Felix}
F.~Schlüter et al.
\newblock \emph{PoS(ICRC2025)835}, 2025.

\bibitem[Scholten et~al.(2024)]{Scholten:2024upn}
O.~Scholten et~al.
\newblock \emph{Phys. Rev. D}, 110\penalty0 (10):\penalty0 103036, 2024.
\newblock \doi{10.1103/PhysRevD.110.103036}.

\bibitem[Schoorlemmer and Carvalho(2021)]{Schoorlemmer:2020low}
H.~Schoorlemmer and W.~R. Carvalho.
\newblock \emph{Eur. Phys. J. C}, 81\penalty0 (12):\penalty0 1120, 2021.
\newblock \doi{10.1140/epjc/s10052-021-09925-9}.

\bibitem[Straub et~al.(2025)Straub, En{\ss}lin, Erdmann, Frank, and
  Zingler]{Straub:2025lsd}
M.~Straub et al.
\newblock \emph{Submitted}, 7 2025.
\newblock URL \url{https://arxiv.org/abs/2507.20555}.

\bibitem[{Tibet AS{\ensuremath{\gamma}}
  Collaboration}(2021)]{TibetASG:2021NatA}
{Tibet AS{\ensuremath{\gamma}} Collaboration}.
\newblock \emph{Nature Astronomy}, 5:\penalty0 460--464, Jan. 2021.
\newblock \doi{10.1038/s41550-020-01294-9}.

\bibitem[van Haarlem et~al.(2013)]{LOFAR:2013jil}
M.~P. van Haarlem et~al.
\newblock \emph{Astron. Astrophys.}, 556:\penalty0 A2, 2013.
\newblock \doi{10.1051/0004-6361/201220873}.

\bibitem[Watanabe et~al.(2025)]{Watanabe:2025njo}
K.~Watanabe et~al.
\newblock \emph{PoS}, ICRC2025:\penalty0 436, 2025.

\end{thebibliography}

\end{document}